\documentclass[conference]{IEEEtran}
\IEEEoverridecommandlockouts

\usepackage{graphicx}
\usepackage{balance}  

\usepackage{graphicx}
\usepackage{url}
\usepackage{syntax}
\usepackage{newfloat}
\usepackage{multicol}
\usepackage{hyperref}

\newcommand{\Comment}[1]{}

\newcommand{\SmallSpace}{\vspace*{-1.5ex}}

\newcommand{\para}[1]{\smallskip\noindent{\bf {#1}.}}

\DeclareFloatingEnvironment[
  fileext   = logr,
  listname  = {List of Grammars},
  name      = Grammar,
  placement = !htbp
]{BNF}

\newcommand{\distance}{5pt}
\usepackage{acronym}
\usepackage[ruled,noend]{algorithm2e}
\usepackage{mathtools}
\usepackage{url}
\usepackage{multirow}
\usepackage{mdwlist}
\usepackage{xspace}
\usepackage{amsmath}
\usepackage{amssymb}
\usepackage{misc}
\usepackage{subcaption}
\usepackage{listings}
\usepackage{enumerate}

\usepackage{enumitem}
\usepackage{adjustbox}
\usepackage{mdframed}

\usepackage{xcolor}

\usepackage[capitalize,nameinlink]{cleveref}
\hypersetup{%
	bookmarksnumbered, bookmarksopen=true, bookmarksopenlevel=1,%
}

\definecolor{mygreen}{rgb}{0,0.6,0}
\definecolor{mygray}{rgb}{0.5,0.5,0.5}

\crefname{figure}{Figure}{Figures}
\crefname{listing}{Query}{Queries}
\crefname{section}{Section}{Sections}
\crefname{table}{Table}{Tables}
\crefname{BNF}{Grammar}{Grammars}
\crefname{algorithm}{Algorithm}{Algorithms}

\SetKwInput{KwInput}{Input}
\SetKwInput{KwOutput}{Output}
\SetKwProg{Fn}{Function}{}{end}

\SetCommentSty{mycommfont}

\newcommand{\incode}[1]{\lstinline{#1}}

\newcommand{\dsl}{\textsc{Saql}\xspace}
\newcommand{\eat}[1]{}

\newcommand{\eg}{{\it e.g.,}\xspace}

\newcommand{\ie}{{\it i.e.,}\xspace}

\usepackage[labelfont=bf,skip=0pt]{caption}
\newcommand{\dist}{2pt}
\acrodef{CEP}{Complex  Event  Processing}
\acrodef{SAQL}{\emph{Stream-based Anomaly Query Language}}

\acrodef{APT}{Advanced Persistent Threats}
\acrodef{SIEM}{Security Information and Event Management}
\acrodef{IDS}{Intrusion Detection Systems}
\acrodef{DSL}{Domain Specific Language}

\eat{
\fancyhf{} 
\fancyhead[C]{Anonymous Submision \#9999 to ACM CCS 2017} 
\fancyfoot[C]{\thepage}

\setcopyright{none} 
\acmConference[Anonymous Submission to ACM CCS 2017]{ACM Conference on Computer and Communications Security}{Due 19 May 2017}{Dallas, Texas}
\acmYear{2017}

\settopmatter{printacmref=false, printccs=true, printfolios=true} 

}

\def\BibTeX{{\rm B\kern-.05em{\sc i\kern-.025em b}\kern-.08em
    T\kern-.1667em\lower.7ex\hbox{E}\kern-.125emX}}

\begin{document}
\date{}

\title{Querying Streaming System Monitoring Data for Enterprise System Anomaly Detection}


%
%

\author{
Peng Gao$^1$\; Xusheng Xiao$^2$\; Ding Li$^3$\; Kangkook Jee$^4$\;  Haifeng Chen$^3$\; Sanjeev R. Kulkarni$^5$\; Prateek Mittal$^5$
	\\
      {\it \small $^1$UC Berkeley\; $^2$Case Western Reserve University\; $^3$NEC Labs America\; $^4$UT Dallas\; $^5$Princeton University}
 	\\
       {\centering \it \footnotesize penggao@berkeley.edu\; xusheng.xiao@case.edu\; \{dingli,haifeng\}@nec-labs.com\; kangkook.jee@utdallas.edu\; \{kulkarni,pmittal\}@princeton.edu}
}

\maketitle

\thispagestyle{empty}
\pagestyle{empty}

\begin{abstract}
The need for countering Advanced Persistent Threat (APT) attacks
has led to the solutions that ubiquitously monitor system activities in each enterprise host, and perform timely abnormal system behavior detection over the stream of monitoring data.
However, existing stream-based solutions lack explicit language constructs for expressing anomaly models that capture abnormal system behaviors, thus facing challenges in incorporating \emph{expert knowledge} to perform \emph{timely anomaly detection} over the large-scale monitoring data.
To address these limitations, we build \dsl, a novel stream-based query system that takes as input, a real-time event feed aggregated from multiple hosts in an enterprise, and provides an anomaly query engine that queries the event feed to identify abnormal behaviors based on the specified anomaly models.
\dsl provides a domain-specific query language, \emph{Stream-based Anomaly Query Language (\dsl)}, that uniquely integrates critical primitives for expressing major types of anomaly models.
%
%
%
In the demo, we aim to show the complete usage scenario of \dsl by (1) performing an APT attack in a controlled environment, and (2) using \dsl to detect the abnormal behaviors in real time by querying the collected stream of system monitoring data that contains the attack traces.
The audience will have the option to 
interact with the system and detect the attack footprints 
in real time 
via issuing queries and checking the 
query
results through a command-line UI.

\end{abstract}

\section{Introduction}

Advanced cyber attacks and data breaches plague even the most protected businesses~\cite{target, equifax}. 
Similar attacks, especially in the form of Advanced Persistent Threats (APTs), are being commonly observed.
These attacks consist of \emph{a sequence of steps} across \emph{many hosts} that exploit different types of vulnerabilities to compromise security.
To counter these attacks, approaches based on \emph{ubiquitous system monitoring} have emerged as an important solution for \emph{monitoring system activities and actively detecting possible abnormal system behaviors}~\cite{backtracking,reduction,gao2018aiql,gao2018saql}. 
System monitoring observes system calls at the kernel level to collect system-level events
that record interactions among system entities (\eg processes, files, and network connections).
Collection of system monitoring data enables security analysts to detect abnormal system behaviors by \emph{continuously searching for anomalies from the streaming data}~\cite{anomalysurvey,idsbook}.

\begin{figure*}
	\centering
	\includegraphics[width=0.825\textwidth]{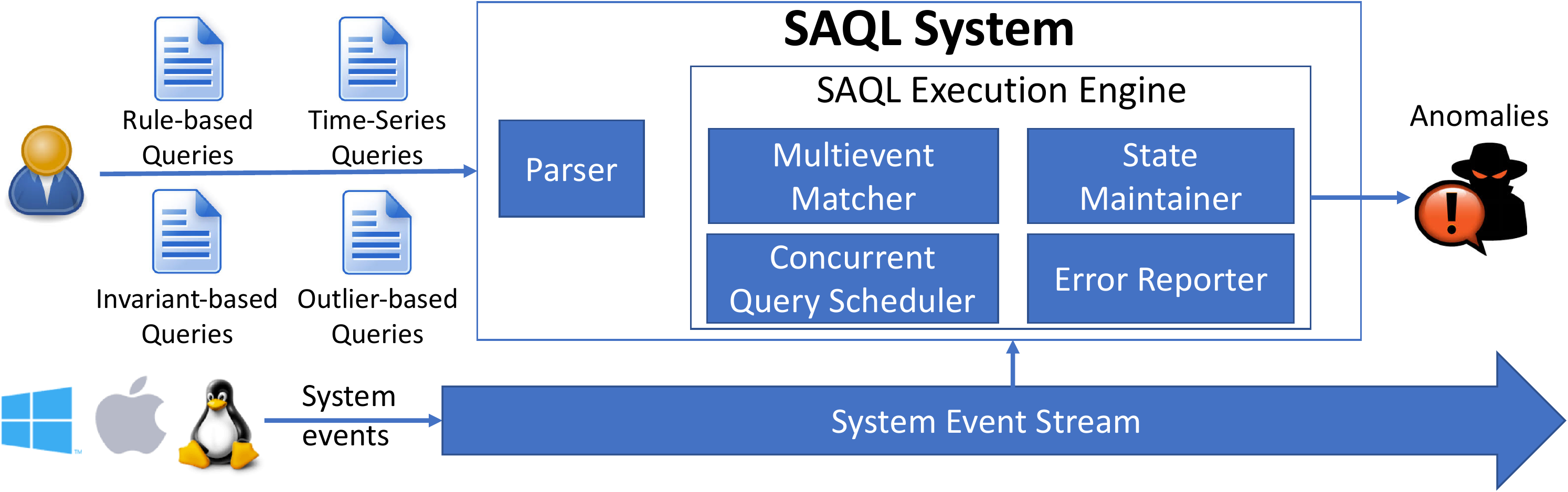}
	\caption{The architecture of \dsl}
	\label{fig:arch}
\end{figure*}

Fighting against advanced attacks such as APTs is a time-critical mission. As such, there is a strong need for a \emph{real-time} anomaly detection system that can find a ``needle in a haystack'' from system monitoring data for preventing additional damage and performing system recovery. However, there are two major challenges for building such system to support effective and timely anomaly detection:
(1) \emph{Expert Knowledge Incorporation}:
Advanced attacks 
typically involve multiple steps exploiting various types of vulnerabilities. Besides, models derived from data have been increasingly used in detecting various types of abnormal behaviors~\cite{idsbook}.
System administrators, security analysts, and data scientists have extensive domain knowledge about the enterprise, including the expected system behaviors.
Fighting against such attacks 
requires the system to provide a unified interface for expressing a broad range of anomaly models while \emph{incorporating the domain knowledge from experts};
(2) \emph{Timely Big-Data Analytics}:
System monitoring produces huge amount of daily logs ($\sim$50GB for 100 hosts per day)~\cite{backtracking,reduction}. 
This requires the system to provide \emph{efficient} real-time data analytics.

Unfortunately, none of the existing 
systems~\cite{cql,streamanomaly1,siddhi,flink} provide a comprehensive solution 
that addresses both of these inherent challenges.
Existing anomaly detection systems focus on building models for specific anomalies based on 
extracted features, rather than providing a unified interface for expressing a broad range of anomaly models via incorporating the expert knowledge.
Existing stream-based query systems 
are designed to work with general-purpose data streams,
and lack explicit language constructs for expressing various anomaly models for our particular problem domain.
Furthermore, to support multiple concurrent queries that access different attributes of the data, these systems have to make multiple copies of the data for the queries, 
and thus is not efficient in handling the big data collected from system monitoring.

To address these challenges, we build \dsl~\cite{gao2018saql}, a stream-based query system that enables security analysts to perform real-time abnormal system behavior detection via querying the stream of system monitoring data.
\dsl takes as input a real-time event feed aggregated from multiple hosts in an enterprise, and provides an anomaly query engine that queries the event feed to identify abnormal system behaviors based on the specified anomaly models.
To facilitate the task of expert knowledge incorporation, \dsl provides a domain-specific query language, \emph{Stream-based Anomaly Query Language (\dsl)}, that uniquely integrates a series of critical primitives for expressing a broad range of anomaly models. 
In particular, \dsl provides
(1) the syntax of event patterns
for specifying relevant system activities and their relationships, which facilitates the specification of \emph{rule-based anomaly models};
(2) the constructs for \emph{sliding windows} and \emph{stateful computation}
that allow stateful anomaly models to be computed in each sliding window over the data stream.
These constructs facilitate the specification of \emph{time-series anomaly models}, \emph{invariant-based anomaly models},
and \emph{outlier-based anomaly models},
which lack support from existing stream-based query systems.
%
To facilitate the task of efficiently handling concurrent queries, \dsl employs a master-dependent-query scheme that groups semantically compatible queries to minimize the data copies of the stream.
Our anomaly query engine leverages the domain-specific characteristics of the system monitoring data and the semantics of the queries to efficiently schedule the execution.


We have deployed the \dsl system in NEC Labs America comprising 150 hosts and made a demo video~\cite{saql-demo-video}.
In our demo, we aim to show the complete usage scenario of \dsl.
To achieve this goal, we perform an APT attack in a controlled environment (for protecting the normal business) that exploits multiple vulnerabilities in the system and exfiltrates sensitive data from database server.
The system monitoring data that contains the attack traces is collected by our data collection agents and sent to the central server, forming an event stream.
We constructed a set of \dsl queries in advanced for detecting the attack behaviors and deployed them over the stream. 
These \dsl queries will continuously 
monitor the stream and report the alerts in real time as we perform the attack.
To easily reproduce the attack data for showcasing different queries, we additionally store the data in databases and have built a stream replayer to replay the data from databases as a data stream.
The audience will have the option to 
interact with the system and detect the attack footprints
via issuing queries and checking the query results through a command-line UI.

\eat{
In our demo,
we asked white hat hackers to perform an APT attack in our deployed environment. The APT attack exploits multiple vulnerabilities in the system
and exfiltrates sensitive data from database server.
The audience will have the option to interact with the system and detect the attack footprints via issuing queries and checking the results through a command-line UI.
}

\section{The SAQL System Architecture}
\label{sec:overview}

 \cref{fig:arch} shows the architecture of the \dsl system.
 \dsl takes an input query from the user 
that specifies certain attack behaviors to be detected, executes the query by checking the specified behaviors against the system event stream, and reports the detection alerts once there exist matches.

\subsection{Data Collection}
\label{subsec:data-collection}


\eat{
System monitoring collects system auditing events about system calls that are crucial in security analysis, describing the interactions among system entities.
As shown in existing studies~\cite{backtracking,reduction,gao2018saql,gao2018aiql}, on mainstream operating systems (Windows, Linux, and Mac OS), system entities in most cases are files, processes, and network connections,
and the collected system calls are mapped to three major types of system events:
(i) file access, 
(ii) processes creation and destruction, and
(iii) network access.
As such, we consider \emph{system entities} as \emph{files}, \emph{processes}, and \emph{network connections}.
We consider a \emph{system event} as the interaction between two system entities represented as \emph{$\langle$subject, operation, object$\rangle$}. 
Subjects are processes originating from software applications (\eg Chrome), and objects can be files, processes, and network connections. 
We categorize system events into three types according to the types of their object entities, namely \emph{file events}, \emph{process events}, and \emph{network events}.
}

System monitoring records kernel-level interactions
among 
system entities as system events.
Each of the recorded event occurs on a particular host at a particular time, thus exhibiting strong spatial and temporal properties.
Following the established convention~\cite{backtracking,reduction,gao2018saql,gao2018aiql},
in our data model, we consider \emph{system entities} as files, processes, and network connections.
We consider a \emph{system event} as the interaction between two system entities represented 
as \emph{$\langle$subject, operation, object$\rangle$} (SVO).
Subjects are processes originating from software applications (\eg Firefox), and objects can be files, processes, and network connections.
We categorize system events into three types according to their objects, namely \emph{file events}, \emph{process events}, and \emph{network events}.

We build data collection agents based on mature system monitoring frameworks: auditd for Linux, ETW for Windows, and DTrace for MacOS. 
Our agents are deployed across servers, desktops, and laptops in the enterprise to collect system auditing events from kernels.
The collected events with critical security-related attributes (e.g., file name, process executable name, PID, IP, port; more details in~\cite{gao2018saql}) are sent to the central server, forming an event stream.

\subsection{SAQL Query Language}
\label{subsec:language}

We build the \dsl language using ANTLR 4. Our language uniquely integrates a series of critical primitives for concisely expressing four major types of anomaly models.

\subsubsection{Rule-based Anomaly Model}
\dsl provides explicit constructs to specify system entities/events, attribute constraints, and event temporal/attribute relationships. This facilitates the specification of rule-based anomaly models to detect known attack behaviors or enforce enterprise-wide security policies.
\cref{case:c5:comp} shows a \dsl query that detects the data exfiltration from database server:
the attacker leverages OSQL utility (\incode{osql.exe}) to dump the database content (\incode{backup1.dmp}) and then runs a malware (\incode{sbblv.exe}) to send the dump back to his host (\incode{XXX.129}).
Four event patterns are declared (Lines 2-5) with a global constraints (Line 1), a temporal relationship (Line 6), and an implicit attribute relationship (Lines 3-4 specify the same \incode{f1} in both events).
Desired attributes of matched events are returned (Line 7) with context-aware syntax shortcuts adopted
(\ie \incode{p1} $\rightarrow$ \incode{p1.exe_name}).

\begin{lstlisting}[captionpos=b, caption={A rule-based \dsl query}, label={case:c5:comp}]
agentid = xxx // SQL database server (obfuscated)
proc p1["%cmd.exe"] start proc p2["%osql.exe"] as evt1
proc p3["%sqlservr.exe"] write file f1["%backup1.dmp"] as evt2
proc p4["%sbblv.exe"] read file f1 as evt3
proc p4 read || write ip i1[dstip="XXX.129"] as evt4
with evt1 -> evt2 -> evt3 -> evt4
return distinct p1, p2, p3, f1, p4, i1 // p1 -> p1.exe_name, i1 -> i1.dstip, f1 -> f1.name
\end{lstlisting}

\eat{
\cref{query:rule} shows a \dsl query for describing an attack step that reads external network (\incode{evt1}), downloads a database cracking tool {\tt gsecdump.exe} (\incode{evt2}), and executes (\incode{evt3}) it to obtain database credentials.
Three event patterns are declared (Lines 1-3). 
It also specifies these events should occur in ascending temporal order (Line 4).

\begin{lstlisting}[captionpos=b, caption={A rule-based \dsl query}, label={query:rule}]
proc p1 read || write ip i1[src_ip != "internal_address"] as evt1
proc p2["%powershell.exe"] write file f1["%gsecdump.exe"] as evt2
proc p3["%cmd.exe"] start proc p4["%gsecdump.exe"] as evt3
with evt1 -> evt2 -> evt3
return p1, i1, p2, f1, p3, p4 // p1 -> p1.exe_name, i1 -> i1.dst_ip, f1 -> f1.name
\end{lstlisting}
}

\subsubsection{Time-Series Anomaly Model}
\dsl provides explicit constructs for sliding windows and stateful computation that allow stateful anomaly models to be computed in each sliding window over the stream.
These constructs lay the foundation for specifying advanced anomaly models (\ie time-series anomaly models, invariant-based anomaly models), which lack support from existing stream-based query systems~\cite{cql, streamanomaly1,siddhi,flink}.
\cref{query:frequency} shows a \dsl query that specifies a time-series anomaly model to 
monitor the network usage of each application and raises an alert when the network usage is abnormally high.
It specifies a 10-minute sliding window (Line 1), collects the amount of data sent through network within each window (Lines 2-4),
and computes the moving average to detect spikes of network data transfers (Line 5). 
In the query, \incode{ss[0]} represents the state of the current window while \incode{ss[1]} and \incode{ss[2]} represent the states of the two past windows respectively (\incode{ss[2]} occurs earlier than \incode{ss[1]}).

\begin{lstlisting}[captionpos=b, caption={A time-series \dsl query}, label={query:frequency}]
proc p write ip i as evt #time(10 min)
state[3] ss {
	avg_amount := avg(evt.amount)
} group by p
alert (ss[0].avg_amount > (ss[0].avg_amount + ss[1].avg_amount + ss[2].avg_amount) / 3) && (ss[0].avg_amount > 10000)
return p, ss[0].avg_amount, ss[1].avg_amount, ss[2].avg_amount
\end{lstlisting}

\subsubsection{Invariant-based Anomaly Model}
\dsl provides explicit constructs for learning invariants of system behaviors under normal operations and using the learned invariants to detect later violations. 
This facilitates the specification of invariant-based anomaly models.
\cref{query:invariant} shows a \dsl query that specifies a 10-second sliding window (Line 1), maintains a set of child processes spawned by the Apache process (Lines 2-4), uses the first ten windows to train the invariant (Lines 5-8), and detects unseen child processes spawned by Apache (Line 9).
\eat{
The specified invariant
represents the set of
child processes spawned by the Apache process in the training stage.
During the 
detection phase, this query generates alerts when an unseen process is forked.
}

\begin{lstlisting}[captionpos=b, caption={An invariant-based \dsl query}, label={query:invariant}]
proc p1["%apache.exe"] start proc p2 as evt #time(10 s)
state ss {
	set_proc := set(p2.exe_name)
} group by p1
invariant[10][offline] {
	a := empty_set // invariant init
	a = a union ss.set_proc //invariant update
}
alert |ss.set_proc diff a| > 0
return p1, ss.set_proc
\end{lstlisting}

\subsubsection{Outlier-based Anomaly Model}
\dsl provides explicit constructs for grouping system behaviors together to detect outliers.
This facilitates the specification of outlier-based anomaly models.
\cref{query:cluster} shows a \dsl query that specifies a 10-minute sliding window (Line 2), computes the amount of data sent through network by the \incode{sqlservr.exe} process for each outgoing IP address (Lines 3-5), and identifies the outliers using DBSCAN clustering (Lines 6-7) to detect the suspicious IP that triggers the database dump.
Note that Line 6 specifies which information of the state forms a comparison point and how the ``distance'' among these points should be computed (\incode{"ed"} represents Euclidean Distance).
%
%

\begin{lstlisting}[captionpos=b, caption={An outlier-based \dsl query}, label={query:cluster}]
agentid = xxx // SQL database server (obfuscated)
proc p["%sqlservr.exe"] read || write ip i as evt #time(10 min)
state ss {
	amt := sum(evt.amount)
} group by i.dstip
cluster(points=all(ss.amt), distance="ed", method="DBSCAN(100000, 5)")
alert cluster.outlier && ss.amt > 1000000
return i.dstip, ss.amt
\end{lstlisting}

\eat{
In addition to detecting outliers through clustering, \dsl also supports the detection through aggregation comparison. 
For example, in \cref{query:cluster}, replacing the \incode{alert} statement with \incode{alert ss.amt>1.5*iqr(all(ss.amt))+q3(all(ss.amt))} gives interquartile range (IQR)-based outlier detection~\cite{casella2002statistical}, and replacing the \incode{alert} statement with \incode{alert ss.amt>3*stddev(all(ss.amt))+avg(all(ss.amt))} gives 3-sigma-based outlier detection~\cite{casella2002statistical}.
}


\subsection{SAQL Query Execution Engine}
\label{subsec:engine}

We build the \dsl system upon Siddhi CEP~\cite{siddhi} so that our system can leverage Siddhi's mature mechanisms to manage the event stream. 
Given an input \dsl query, the multievent matcher matches the events in the stream against the event patterns specified in the query. 
If the query involves stateful computation, the state maintainer maintains the states of each sliding window computed from the matched events.
To efficiently handle the execution of multiple concurrent queries, the concurrent query scheduler employs a master-dependent-query scheme.
In the scheme, concurrent queries are divided into groups based their semantic compatibilities, with each group having a master query and several dependent queries. 
The scheme enforces that the queries in a group will share a single copy of the stream data for execution.
Only master queries have direct access to the data stream, and the execution of the dependent queries leverages the intermediate execution results of their master query. 
In this way, unnecessary data copies of the stream can be significantly reduced.
The error reporter reports the errors during the query execution.
Alerts are generated when the alert conditions specified in the query are matched by the event stream.

\begin{figure}[t]
	\centering
	\includegraphics[width=0.45\textwidth]{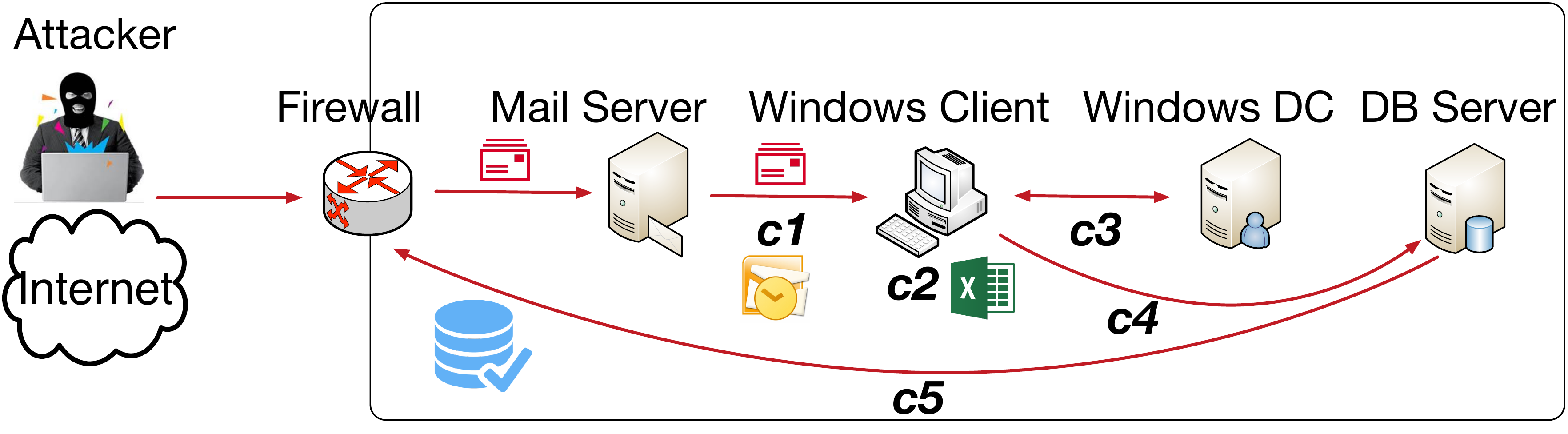}
	\caption{Demonstration setup for the APT attack}
	\label{fig:outlook-apt}
\end{figure}

\begin{figure}[t]
	\centering
	\includegraphics[width=0.45\textwidth]{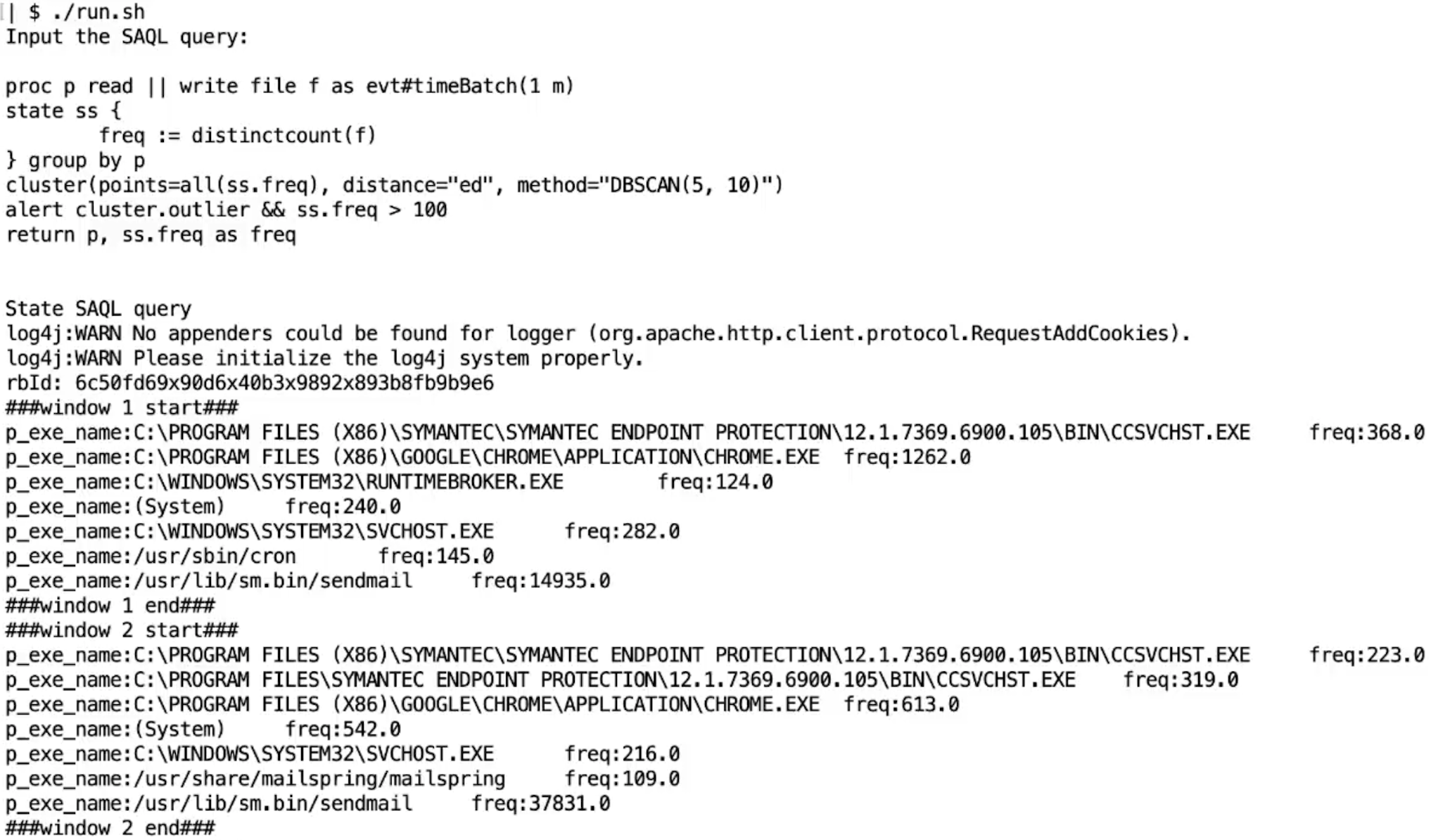}
	\caption{Command-line UI of the \dsl system}
	\label{fig:ui}
\end{figure}

\section{Demonstration Outline}
\label{sec:demo}

\para{Demonstration Setup for The APT Attack}
We deployed \dsl in NEC Labs America comprising 150 hosts.
%
%
The purpose of our demo is to illustrate the complete usage scenario of \dsl and showcase its superiority in enabling timely abnormal system behavior detection. 
To achieve this goal, 
we perform an APT attack in a controlled environment  (\cref{fig:outlook-apt}) using known exploits. The APT attack consists of five steps as follows:


\begin{itemize}[label={\arabic*.}, noitemsep, topsep=1pt, partopsep=1pt, listparindent=\parindent, leftmargin=*]
	\item[\emph{c1}] \emph{Initial Compromise}: The attacker sends a crafted email to the victim. The email contains an Excel file with a malicious macro embedded. 
	
	\item[\emph{c2}] \emph{Malware Infection}: The victim opens the Excel file through the Outlook client and runs the macro, which downloads and executes a malicious script (CVE-2008-0081~\cite{cveexcel}) to open a backdoor for the attacker.
	
	\item[\emph{c3}] \emph{Privilege Escalation}: The attacker enters the victim's machine through the backdoor, scans the network ports to discover the IP address of the database, and runs the database cracking tool ({\tt gsecdump.exe}) to steal the credentials of the database.
	
	\item[\emph{c4}] \emph{Penetration into Database Server}: Using the credentials, the attacker penetrates into the database server and delivers a VBScript to drop another malicious script, which creates another backdoor.
	
	\item[\emph{c5}] \emph{Data Exfiltration}: With the access to the database server, the attacker dumps the database content using {\tt osql.exe} and sends the data dump back to his host.
\end{itemize}

\para{Construction of SAQL Queries}
We constructed 8 \dsl queries (more details in~\cite{saql-demo-queries}) in advance for detecting the 
attack behaviors.
For each attack step, we construct a rule-based \dsl query by leveraging the knowledge of the attack.
Furthermore, we construct three advanced anomaly queries, assuming no knowledge of the attack details:

\begin{itemize}[noitemsep, topsep=1pt, partopsep=1pt, listparindent=\parindent, leftmargin=*]
	\item We construct an invariant-based anomaly query to detect the scenario where Excel executes a malicious script that it has never executed before (\ie step \emph{c2}):
	The invariant contains all unique processes started by Excel in the first 100 sliding windows. 
	New processes that deviate from the invariant are reported as alerts. 
	
	\item We construct a time-series anomaly query based on SMA to detect 
the scenario where abnormally high volumes of data are exchanged via network on the database server (\ie step \emph{c5}): 
	For every process on the database server, this query detects the processes that transfer abnormally high volumes of data to the network. 
	
	\item We construct an outlier-based anomaly query to detect processes that transfer high volumes of data to the network (\ie step \emph{c5}): 
	The query detects such processes through peer comparison based on DBSCAN.
\end{itemize}

\para{Demonstration Procedure}
We start the demonstration by constructing and executing the \dsl queries using a command-line UI (\cref{fig:ui}).
As we perform the attack, the \dsl queries will continuously monitor the stream and report the alerts when the abnormal attack behaviors are detected.

\begin{figure}[t]
	\centering
	\includegraphics[width=0.45\textwidth]{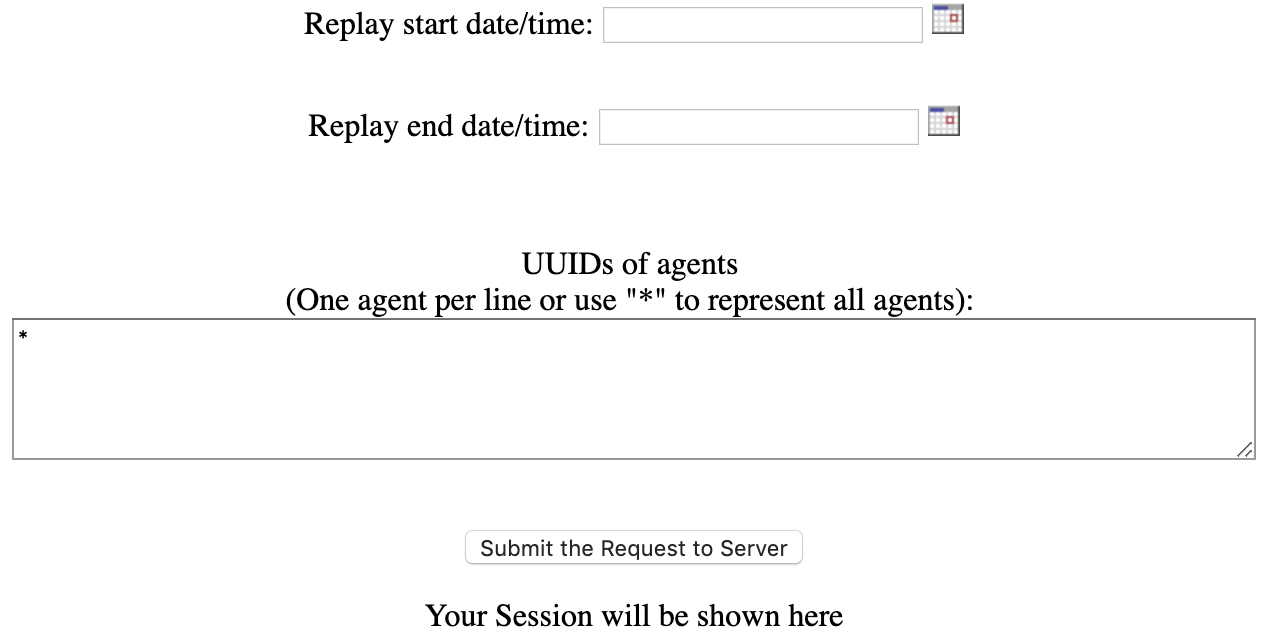}
	\caption{Stream replayer}
	\label{fig:replayer}
\end{figure}

To easily reproduce the streaming attack data for showcasing different queries, we additionally store the data in databases and have built a stream replayer to replay the data from databases as a data stream.
Our stream replayer has a web-based UI (\cref{fig:replayer}) that lets us choose the hosts and the start/end time to replay the system monitoring data.

\eat{
In our demo, we will prepare some scripts to let the audience \emph{leverage existing exploits to perform the APT attack under our guidance}, and detect the attack footprints in real time via issuing \dsl queries and checking the reported alerts.
}


\section{Conclusion}

We have presented \dsl, a novel system for detecting abnormal system behaviors in enterprises via querying streaming system monitoring data.
\dsl provides an expressive domain-specific language to express a wide range of anomaly models.

\para{Acknowledgement}
This work was supported in part by DARPA
N66001-15-C-4066.
Any opinions, findings, and conclusions made in this material are those of the authors and do not necessarily reflect the views of the funding agencies.


\balance

\bibliographystyle{abbrv}
\bibliography{ref} 


\end{document}